\documentclass[a4paper,fleqn,12pt,twoside]{article}
\usepackage[headings]{espcrc1}

\readRCS
$Id: espcrc1.tex,v 1.2 2004/02/24 11:22:11 spepping Exp $
\usepackage{graphicx}

\newcommand{\AmS}{{\protect\the\textfont2
  A\kern-.1667em\lower.5ex\hbox{M}\kern-.125emS}}

\title{QCD Effective action at high
temperature and small chemical potential}

\author{C. Villavicencio and E. S. Fraga\\
~\\
Instituto de F\'{\i}sica, Universidade Federal do Rio de Janeiro, C.P. 68528,
Rio de Janeiro, RJ 21941-972, Brazil.}

\runtitle{QCD effective action at high
temperature and small chemical potential}
\runauthor{C. Villavicencio and E. S. Fraga}

\usepackage{slashed}
\newcommand{\ttin}[1]{{\mbox{\tiny{#1}}}}
\begin{document}

\maketitle

\bigskip
\begin{abstract}
\noindent
We present a construction of an effective Yang-Mills action for QCD, from
the expansion of the fermionic determinant in terms of powers of the chemical
potential at high temperature for the case of massless quarks.
We analyze this expansion in the perturbative region and find
that it gives extra spurious information. We
propose for the non-perturbative sector a simplified effective action
which, in principle, contains only the relevant information.
\end{abstract}

\section{Introduction}

\noindent
There has been an increasing interest in the few last years in the
\emph{sign problem} or \emph{phase problem} in QCD.
For finite chemical potential, the fermion determinant matrix is
non-positive
definite, so it is not possible to perform Monte-Carlo simulations
\cite{Hands:2001jn,Laermann:2003cv}.
The Glasgow method \cite{Barbour:1991vs} and
 rewheighting techniques \cite{Fodor:2001au} have made great advances in
the description of phase transitions on the lattice, considering  a set of
parameters near the transition line. However, the problem is still far from
being  solved. 

There is special interest in the region of high temperature and low
chemical potential, since it corresponds to the high-energy heavy-ion collision
scenario. 
In this regime it is possible to expand the fermion
determinant in powers of $\mu/T$
\cite{Gottlieb:1987ac,Allton:2003vx,Allton:2005gk}.
This kind of expansion is what we are interested in investigating. 
As one performs this expansion, some
questions arrise: what order is enough to cut the
expansion?
 which is its range of validity?
how to distinguish
between  real and complex terms?
   Differently from what is done in the previously  mentioned papers, we keep
the
expanded series as a functional of the gauge fields, giving rise to an
effective non-local Yang-Mills action.

\section{Basic idea of the expansion}

\noindent
For simplicity, consider one-flavor massless quarks. 
The QCD generating
functional at finite chemical potential in euclidean space is defined
as
\begin{equation}
 {\cal Z}=\int {\cal D}G
\det(-i\slashed{D}+i\mu\gamma_4)e^{-S_{\mbox{\tiny YM}}[G]},
\end{equation}
where $G$ are the gluon fields, also present in the covariant
derivative $D$,
and $S_{\mbox{\tiny YM}}$ is the Yang-Mills action. 
We can expand the fermion determinant in powers of the chemical
potential
assuming that $\mu<\Lambda\sim T$:
\newpage
\begin{eqnarray}
&& 
\hspace{-.65cm}\det (-i\slashed{D}+i\mu\gamma_4)\nonumber\\
&&\hspace{.5cm} =\det
(-i\slashed{D})
\exp\Bigg\{-\sum_{s=1}^\infty\frac{(i\mu)^s}{s}
\int_{Y_1\cdots Y_s}   
\mbox{Tr}~\gamma_4S(Y_2,Y_1)\gamma_4S(Y_3,Y_2)\dots
\gamma_4S(Y_1,Y_n)\Bigg\},
\end{eqnarray}
where $S(Y_b,Y_a)$ is the dressed fermion propagator, which can be
expressed as
a series in powers of the gauge field and the free fermion
propagator. 
The expansion, then, will contribute to additional terms in the
Yang-Mills effective action
\begin{equation}
S_{\mbox{\tiny YM}}^{\mbox{\tiny eff}}
=S_{\mbox{\tiny YM}}
+\sum_{n=0}^\infty\int_{X_1\cdots X_n}
{\Gamma^{(n)}}^{\mu_1\dots\mu_n}_{a_1\dots a_n}(X_1,\dots,X_n)
~G^{a_1}_{\mu_1}(X_1)\dots G^{a_n}_{\mu_n}(X_n),
\end{equation}
where the vertices $\Gamma^{(n)}$ are series in powers of the chemical
potential:
$\Gamma^{(n)}=\sum_{s\geq 1}\Gamma^{(n,s)}$, with $\Gamma^{(n,s)}\sim
\mu^s$.
In this way we have a positive-definite fermion determinant and the
contribution from the chemical potential will be part of an effective
Yang-Mills action.
The problem now is to find criteria to cut this infinite series in
terms of the chemical potential and gauge fields.

\section{ Perturbative QCD}

To test the expansion, we study it firstly in the perturbative
sector. 
Setting $G\to gG$, we find naturally the way to cut the gauge field series,
which is given by the order of perturbative corrections.
For one-loop corrections, we just need the effective action up to order $g^2$,
so
\begin{equation}
S_{\ttin{YM}}^{\ttin{eff}}=S_\ttin{YM}+\Gamma^{(0)}+g^2\int_{XY}
{\Gamma^{(2)}}_{\mu\nu}^{ab}(X,Y)G_\mu^a(X)G_\nu^b(Y).
\end{equation}
We calculate, then, the pressure up to order $g^2$.
Considering the expansion on chemical potential up to order $\mu^4$, we
surprisingly obtain the same result as in the usual pertubative QCD
calculation with the chemical potential included in the quark
propagator \cite{Kapusta:1989tk}.

However, the next terms in the chemical potential series are non-zero. In
particular, the next vacuum contribution for the pressure is
\begin{equation}
P^{(6)}_0=-\frac{\Gamma^{(0,6)}}{\beta V}
\approx 0.9334 N_c\frac{\mu^6}{\pi^2T^2}~,
\end{equation}
 which is negligible only for $\mu \ll \Lambda$. More than
establishing the range of validity, this result shows that this kind
of expansion contains some spurious information, at least for the massless
case, and in principle the only indication on where to cut the
 series is the dimension of the desired observable.

\section{ Non-perturbative QCD}

For the  non-perturbative regime, we need another criterium to cut
the
series.
The most appropriate is the Weinberg power counting
\cite{Weinberg:1978kz}, i.e. 
considering the external
momentum as a small parameter, $p< \Lambda$, and also assuming that
the
chemical
potential and the expectation value of the fields are of the same order,
$\mu\sim G\sim p$.
This gives a low-energy theory where the effective Lagrangian can be
expanded in soft modes.
In the case of Yang-Mills theories, this soft-mode expansions for
high temperatures correspond to the Hard Thermal Loop
(HTL) approximation \cite{Braaten:1991gm,Frenkel:1991ts}.
The minimal action  must be of order $p^4$. 
So,  applying the power
counting, the minimal effective action is
\begin{equation}
S_\ttin{YM}^\ttin{min}[G,\mu]
=S_\ttin{YM}[G]
+S^{(0,2)}(\mu) +S^{(0,4)}(\mu) + S_\ttin{LO}^{(2,2)}[G,\mu]+
S_\ttin{LO}^{(3,1)}[G,\mu]~,
\end{equation}
where the indices $(n,s)$ denote powers of $n$ in the gauge field and
$s$ in the chemical potential, and LO means leading order in the HTL
approximation.
In this case, the appearance of imaginary terms will happen for
 $S_\ttin{LO}^{(3,1)}$.
   
The whole series of gauge fields is gauge invariant in every
order in the $\mu$ expansion, i.e. $\sum_nS^{(n,s)}$ is gauge
invariant for all $s$. In the case of the minimal effective action,
the contributions $S^{(2,2)}$ and $S^{(3,1)}$ are independently gauge
invariant, and  HTL preserves this feature.

Of course this effective action must be tested, but in principle it
could
provide rich information on the confinement-deconfinement transition
line.

\section*{Acknowledgements}
\noindent
We thanks Phillipe de Forcrand for comments and
suggestions.
We acknowledge financial support from   CAPES, CLAF, CNPq, FAPERJ and
FUJB-UFRJ.


\begin{thebibliography}{10}

\bibitem{Hands:2001jn}
S. Hands,
\newblock Nucl. Phys. Proc. Suppl. 106 (2002) 142.

\bibitem{Laermann:2003cv}
E. Laermann and O. Philipsen,
\newblock Ann. Rev. Nucl. Part. Sci. 53 (2003) 163.

\bibitem{Barbour:1991vs}
I.M. Barbour and A.J. Bell,
\newblock Nucl. Phys. B 372 (1992) 385.

\bibitem{Fodor:2001au}
Z. Fodor and S.D. Katz,
\newblock Phys. Lett. B 534 (2002) 87.

\bibitem{Gottlieb:1987ac}
S.A. Gottlieb,
W. Liu, D. Toussaint and R.L. Renken, R.L. Sugar,
\newblock Phys. Rev. Lett. 59 (1987) 2247.

\bibitem{Allton:2003vx}
C.R. Allton,
S. Ejiri, S.J Hands, O. Kaczmarek, F. Karsch, E. Laermann and C. Schmidt,
\newblock Phys. Rev. D 68 (2003) 014507.

\bibitem{Allton:2005gk}
C.R. Allton, M. Doring, S. Ejiri, S.J. Hands, O. Kaczmarek, F. Karsch, E. and
K. Redlich,
\newblock Phys. Rev. D 71 (2005) 054508.

\bibitem{Kapusta:1989tk}
J.I. Kapusta,
\newblock Finite Temperature Field Theory, Cambridge University Press,
1989.

\bibitem{Weinberg:1978kz}
S. Weinberg,
\newblock Physica A 96 (1979) 327.

\bibitem{Braaten:1991gm}
E. Braaten and R.D. Pisarski,
\newblock Phys. Rev. D 45 (1992) 1827.

\bibitem{Frenkel:1991ts}
J. Frenkel and J.C. Taylor,
\newblock Nucl. Phys. B 374 (1992) 156.

\end{thebibliography}

\end{document}